\documentclass[ aps,  jmp, amsmath,amssymb, prb,  preprint,]{revtex4-1}%
\usepackage{amssymb}
\usepackage[ansinew]{inputenc}
\usepackage{graphicx,color,calc}
\usepackage{amsmath}
\usepackage{amsfonts}
\usepackage{graphicx}
\usepackage{epstopdf}%
\providecommand{\U}[1]{\protect\rule{.1in}{.1in}}
\newcommand{\ttref}[2]{}

\begin{document}
\title{Velocity distribution of neutral particles ejected from biological material
under ultra short laser radiation }
\author{Wolfgang Husinsky}
\email{husinsky@iap.tuwien.ac.at}
\author{Hatem Dachraoui}
\altaffiliation[Present address:]{Molecular and Surface Physics, Faculty of Physics, Bielefeld University, Germany}

\affiliation{Institut für Allgemeine Physik, Vienna University of Technology, Wiedner
Hauptstrasse 8-10, A-1040 Wien, Austria}

\begin{abstract}
Neutral particles ejected from biological material under ultra short laser
ablation have been investigated by laser post-ionization time-of-flight mass
spectrometry. It could be shown, that beside ionized species, a substantial
amount of neutral particles is ejected. A temporal study of the ablation plume
is carried out by recording neutral particle time-of-flight mass spectra as a
function of delay time between the ablation and post-ionization pulse. Close
the ablation threshold, the mechanism of ejection is found to be of
predominantely mechanical nature, driven by the relaxation of the
laser-induced pressure. In this regime of stress confinement, the ejection
results in very broad velocity distributions and extremely low velocities.

\begin{description}
\item[PACS numbers] 79.20.Ds, 79.20.Eb, 42.65.Re, 42.65.Sf.

\end{description}
\end{abstract}
\maketitle



\section{Introduction}

The interaction of ultra short laser light with biological tissue is the basis
for an immense and still expanding number of medical applications of lasers.
Tissue processing with ultra short laser pulses has been of growing interest
due to the high precision achieved. Refractive surgery inside the human eye or
'nano-surgery' of single biological cells are striking examples
\cite{Luba2003}. The potential of ultra short laser radiation for surgical
applications can be considered as established \cite{Nie04,Luba04}.

While the nature of the mechanisms behind ultrafast laser ablation of
biological targets has been studied theoretically quite extensively, mostly
using molecular-dynamics (MD) simulations \cite{Leveugle04,Leveugle2004}, the
number of detailed experimental investigations is very limited. For example,
molecular dynamics (MD) ablation studies of organic solids have suggested two
principal classes of ablation mechanisms based on the laser parameters such as
fluences and pulse widths: ($i$) photochemical processes in which the breakup
of the material is the result of strong tensile stresses (spallation)
\cite{Lorazo2006}. ($ii$) photothermal ablation, where laser energy absorption
is followed by its conversion into heat, which finally results in high
temperatures; possible consequences include homogeneous nucleation, phase
explosion and vaporization \cite{Danny02}.

For femtosecond laser pulse irradiation, the exact dynamics of the ablation
process is not well understood as it appears to be a complex combination of
these different mechanisms and also a function of the material properties
which may change during the course of ablation.

In this paper we analyze the mechanisms of material ejection from biological
tissue under ultra short laser radiation, as well as their pulse width
dependence. Using laser postionization mass spectrometry techniques we follow
the temporal evolution of the plume composition. The measured velocity
distributions of the ejected neutral particles are extremely broad and cannot
be described by a Maxwell Boltzmann distribution. For fluences close to the
ablation threshold but below that for plasma formation, laser induced pressure
due to overheating of the irradiated material is identified as the key
processes that determine the dynamics of laser ablation. \bigskip

\section{EXPERIMENTAL SETUP}

The experiments have been performed in an UHV chamber equipped with a
reflectron-type time-of-flight (TOF) mass spectrometer for detecting laser
ablated particles. Laser-Post-Ionization is used for neutral particle
detection. Ablation and post ionization is performed with ultra short laser
radiation by a system which consists of two multipass CPA Ti: Sapphire
amplifiers (Femtopower) seeded from a common mode locked Ti:Sapphire
oscillator (Femtosource). The system, operating at a repetition rate of 1~kHz,
provides laser pulses at a center wavelength around 800~nm (1.5~eV photons)
with a typical duration of 25~fs. The ablation beam was incident on the
surface at an angle of 45 measured from the surface normal and was focused to
a spot size of typically 100-500~$%
\mu
$m in diameter. The ablation- and post-ionizing laser beam can be delayed with
respect to each other (ablation and post-ionizing beam). The post-ionizing
laser beam was crossing 2.5 mm normal in front of the target. The base
pressure in the UHV chamber was approximately 10$^{-7}$ mbar, since baking the
chamber is not possible with biological samples. In addition, one of the two
amplifiers was equipped with an opto-acoustic dazzler, which allowed pulse
shaping of the laser beam. This measurement can be achieved despite the many
experimental complexities that result from the not ideal vacuum conditions.
Soft tissues (i.e. cornea), in particular, pose problems in vacuum
environments. We have chosen hard human tissues (bone, tooth) material for our
investigations. The main components of human teeth are enamel and dentin.
Enamel contains 95\% hydroxyl-apatite (Ca10 (PO4)6 (OH)2), 4\% water and 1\%
organic material. Furthermore, it contains impurities like Cl, Na, K, F or Mg.
Dentin contains about 70\% (Ca10 (PO4)6 (OH)2), 20\% organic components
(collagen fibers) and10\% water. Human bone is made up of 50-60\%
hydroxyl-apatite, 15-20\% water, 1\% phosphates (inorganic components), 20\%
collagen and 2\% proteins (organic components). \bigskip

\section{Results and discussion}



Under ultra short laser irradiation, neutrals as well as ions are ejected from
bone- and tooth-material. The detection and measurement of ablated ions is
relatively simple and has been reported previously. The mass spectra of
particles ablated from tooth material and bone material are very similar and
have the basic features in common. Therefore it is sufficient to display and
to discuss the experimental results of one example, and we have chosen the
data of tooth. A typical mass spectrum measured at laser fluence $F=~200%
\operatorname{mJ}%
/%
\operatorname{cm}%
^{2}$ is shown in Figure 1. The dominant signal (highest yield) arises for m/z
= 40, with m ion mass and z the ion charge. In addition, the ion spectrum
shows the presence of many weak peaks attributable to the presence of organic
C$_{m}$H$_{n}$ ions. The observation of relative large particles is a
characteristic signature that photomechanical effects induced by pressure
relaxation play a crucial role in the ablation of biological tissue. Further
analysis was focused on the detection of the neutral particles. Typical mass
spectra of the photoionized neutral particles obtained with fresh and etched
ablated surfaces are shown in Figure 2. The spectra were measured for an
ablation laser fluence of about 200 mJ/cm$^{2}$ and at time delay $\Delta$t
=~5 $%
\mu
$s between the ablation and the post ionization laser pulses. As can easily be
seen, the mass spectrum is significantly modified. In addition to the large
two peaks at$m/z=26$ and $m/z=28$, organics C$_{m}$H$_{n}$ are observed. Only
with the fresh surface, the organics C$_{m}$H$_{n}$ are observed in relatively
high abundance. In contrast to the cation spectrum the neutral mass spectrum
shows no intensity at the mass$/z$ (39, 96, 103 and 112). This observation
suggests that many parent clusters undergo non linear dissociation during the
post-ionisation event. Thus, the neutral spectrum consists essentially of
fragment particles rather than direct ejected particles. Furthermore,
comparison with the ion spectrum reveals, that close to the ablation threshold
neutral and ions are ejected with a majority of neutrals.

Given the large number of clusters found in the plume in this regime, ablation
exhibits a mechanical rather than thermal character.



Laser ablation in the short-pulse regime involves different processes which
are effective at different deposited laser energies. In order to investigate
the mechanisms of the ejection close to the ablation threshold we have
measured the time-of-flight (TOF) distributions of several neutral particles.
Contrary to measurements of the velocity of ejected ions, the measurement of
neutrals is more informative due to the physical ablation process itself,
since neutrals are not influenced by potentials in the surface (the
surface-space-charge effect), which can obscure the results quite
substantially and have led to different interpretations and velocities
reported (eV-keV) \cite{Kreitschitz94}. The measurements are performed with
etched surfaces when no signal change with the number of laser pulses
occurred. Here it is important to note that the system relaxes within a few
hundreds picoseconds after the excitation ($<<~$the laser repetition period).
Figure 3 shows the integration over the 28/z and 54/z peaks as a function of
the flight time between target and ionizing laser. The TOF spectra can be
directly converted into a velocity or energy distribution. Both $m/z=28$ and
$m/z=54$ distribution are maximized at 5~$%
\mu
$s time delay that corresponds to $600~%
\operatorname{m}%
/%
\operatorname{s}%
$ velocity. This velocity of $600~%
\operatorname{m}%
/%
\operatorname{s}%
$, observed for most particles, seem to be rather low at first sight, but
similar slow velocity distributions have been predicted in earlier MD
simulations of laser ablation of organic solids in the stress confinement
irradiation regimes \cite{Zhigilei00}. For all TOF data of biological samples
measured, there is a clear discrepancy between the data and a best fit
according to the Maxwell-Boltzmann (MB) distribution. The experimental TOF
distributions are broader than a MB distribution. For comparison, we also
include in figure 3 the TOF distribution of the emitted neutral Fe (m/z=56)
from a steel target. In contrast to the biological TOF distributions, the
measured Fe distribution can satisfactorily be described by a MB distribution.
As we have clearly demonstrated previously in \cite{Dachraoui06,dachr2,bashir}%
, in the case of a metal, matter removal can be attributed to phase explosion
(photothermal effect). Based on these facts, photothermal effects are, under
these conditions, not responsible for the ejection of material from biological tissue.


For longer time delays $\Delta t>~20~\mu%
\operatorname{s}%
$, neutral particles are observed in the mass spectra. Moreover within the
experimental repetition period of 1 ms, the TOF distribution not recovers to
the initial state. This feature was confirmed by observing neutral particles
in the mass spectra recorded at negative delay time. Negative time delay
corresponds to the post-ionization pulse preceding the ablation pulse. These
extremely slow particles explain the observed background in the TOF
distributions. This brings us to an obvious question: what is the origin of
these particles?

It is evident from the low temperature of the ejected particles that the
physical processes leading to material ejection have predominantely mechanical
character. Two observations support photomechanical mechanisms: (i) the
neutral TOF distributions do not follow the MB distribution. (ii) the $28/z$
and $54/z$ peaks show the same temporal evolution; this constitutes a strong
indication that the origin of the particles with low velocities ($600%
\operatorname{m}%
/%
\operatorname{s}%
$) could be due to the ejection of larger clusters, which are fragmented by
the postionization pulse into relatively small C$_{m}$H$_{n}$ fragments. Our
data is consistent with experimental \cite{Itzkan95,Albagli94} and MD
simulation\cite{Zhigilei00,Zhidkov01} results of laser ablation of biological
tissue, where photomechanical effects has been identified as the main
mechanisms responsible for the ablation in the regime of stress confinement.
Furthermore, a plausible explanation of the observed neutrals particles at
time delay around 1 ms is the separation of a large surface layer from the
sample with a velocity of 30 m/s. This behavior is very similar to that
observed in simulations of ablation of organic solids \cite{Zhigilei99}. In
this  theoretical modelling the authors demonstrated that irradiation under
conditions of stress confinement can lead to the ejection of complete layers
of the material.

The origin of the broad distributions can be related primarily to layer
(depth) effects. Confirmation can be found in the literature. Earlier
molecular-dynamics simulations of ablation in organic solids have shown that
different regions form in the sample during the ablation process. These
regions differ in their expansion dynamics and in the pressure relaxation they
follow. This suggests that different ejection conditions for molecules,
depending upon their original depth in the substrate. This means that the
total TOF distribution might be a superposition of Maxwell-Boltzmann
distributions with different stream velocities. Moreover, subsequent
collisions in the expanding plume can lead to further broadening of the
distribution. The experimental TOF distribution ($m/z=28$) can satisfactorily
be described by a shifted Maxwell-Boltzmann distribution \cite{Zhigilei97}.
So, a fit of the TOF distributions ($m/z=28$) to a modified Maxwell-Boltzmann
distribution results in temperatures of $1300%
\operatorname{K}%
$.

We can derive from the data a basic conclusion: the relevance of stress
confinement as an important contribution to the ablation mechanism. In this
scenario, the thermal expansion of the volume heated by a laser pulse
generates compressive stresses. The subsequent propagation of these stresses
from the free surface can transform them into tensile stresses of sufficient
strength to cause mechanical fractures parallel to the surface of the sample
and ejection of the upper layers. The spallation proceeds through nucleation,
growth and coalescence of voids within the spallation region
\cite{Paltauf03,Zhigilei03}.

As already demonstrated in previous experimental and molecular dynamics
simulation studies \cite{Dachraoui06,Lorazo03}, the ablation mechanisms, and
the parameters of the ejected plume have a strong dependence on the laser
pulse duration. Figure 4 shows the measured TOF distribution of the ablated
neutral ($m/z=28$) for  two different additional laser pulse widths
$\Delta\tau=350$ and $500~%
\operatorname{fs}%
$ (in addition to the data for the regular $25%
\operatorname{fs}%
$ pulses). The energy per pulse was kept constant. In this regime of stress
confinement, pulse width variation allows us to obtain important information
about the dependence of the material ejection on the laser peak intensity as a
function of $\Delta\tau$. It is well known that the photomechanical fracture
is determined by the interplay between the tensile pressure and the thermal
softening due to the laser heating: higher tensile pressure (up to -150 MPa)
does not cause mechanical fracture \cite{Danny03,Zhigilei03}.


Much broader TOF distributions with slow velocities are observed in Figure 4a,
which can not be described by a Maxwell-Boltzmann distribution. For a pulse
width $\Delta\tau~=~500%
\operatorname{fs}%
$, the velocity distributions can be roughly described by a modified
Maxwell-Boltzmann distribution. Simulations demonstrate that a broader
distribution is characteristic for the plume ejected in the stress confinement
regime \cite{Zhigilei00}. Furthermore, a comparison with figure 3 (TOF
distribution at $25%
\operatorname{fs}%
$, circles ) reveals two interesting features in this measurement: (i) a clear
shift of the TOF distributions to lower velocities. A similar effect of
temperature-decrease has been observed earlier in MD simulations of laser
ablation of molecular samples. The authors have related the decrease in the
temperature of the plume to enhanced ejection and rapid material
disintegration \cite{Zhigilei00}, even though here a comparison between our
results and the simulations has to be regarded with care (these MD simulations
were performed for 15 and 150 ps duration pulses). We will return to this
point below. (ii) Increase in the background value with increasing the pulse
duration from 25 to 350 fs. This observation reflects the increase in the
abundance of large cluster in the plume.

The dependence of the amount of $m/z=28$ material removed in the time window
of about $1%
\operatorname{ms}%
$ versus laser pulse width is shown in figure 4b. The total ablated yield
increases with increasing the laser pulse width. This observation can explain
the observed decrease in the temperature of the plume (figure 4c). Our
interpretation of the increase in the total amount of the ejected material
takes into account two main effects: (1) An enhancement in the amplitude of
the pressure wave and (2) under these radiation conditions, the balance
between the amplitude of the tensile component and the thermal softening is
more favourable for photomechanical effects (such as spallation).

In the first scenario, in spite of  the dominance of photomechanical ablation
in the first case (short pulse $\Delta\tau=25%
\operatorname{fs}%
$, high peak intensity), one can imagine that this regime is characterised by
the occurrence of photothermal effects in addition to the photomechanical
effects. Because laser-induced spallation can be initiated at energy densities
much lower than those required for phase explosion and vaporization, we
believe that the increase in the pulse duration leads to an irradiation regime
below the thermal ablation threshold and thus phase explosion cannot occur and
account for material ejection in this regime. So, more laser energy is
converted to mechanical energy of the thermoelastic stress wave. This is
probably the reason why spallation becomes more dominant. Second, The etch
depth is determined not only by the amplitude of the tensile component of the
pressure wave, but also the temperature gradient produced within the
irradiated surface region. Close to the ablation threshold a decrease of the
peak intensity leads to an increase in the tensile pressure. This means that
tensile-wave-mediated effects become more dominant for  longer pulses
($350-500%
\operatorname{fs}%
$).

\section{CONCLUSIONS}

In summary, we have investigated the mechanisms of the femtosecond laser
ablation of biological material. At laser fluences above the ablation
threshold, our results indicate that photomechanical effects, driven by the
relaxation of high thermoelastic pressure, are responsible for the collective
material ejection. The velocity distributions are broader than a Boltzmann
distribution and exhibit an offset (extremely slow particles), presumably
fragmentation products of larger clusters or surface layer. A comparison of
the results the velocities obtained with $25$ and $350%
\operatorname{fs}%
$ laser pulses also reveals a number of differences that can have important
implications for practical applications of laser ablation. Larger ablated
volume and broader velocity distributions are produced at the same laser
fluences for pulse longer $\Delta\tau=350%
\operatorname{fs}%
$ than for $\tau=25%
\operatorname{fs}%
$. The experiments presented demonstrate the potential for further
investigations to identify the ablation mechanisms of biological material at
high laser fleunces. Experiments with soft tissues should follow. \bigskip

\section{ACKNOWLEDGMENTS}

This work has been partly supported by the Austrian Science Foundation FW
under project numbers P13756-N02 and P15937-N02. \bigskip

\bibliographystyle{prsty}
\bibliography{Collection}

\newpage

\section{FIGURECAPTIONS}
Fig.1: Typical mass spectrum of ions emitted from tooth sample under ultra short laser radiation (25 fs, 800nm F=~200 mJ/cm$^{2}$).

\bigskip
Fig. 2: Mass spectra of neutral particles of tooth sample produced at F=~200 mJ/cm$^{2}$ and detected at $\Delta$t = 5 $µ$s with relatively fresh (a) and etched (b) surfaces.

\bigskip
Fig. 3: Measured TOF distribution of m/z=28 and 54 neutral particles ablated from tooth material with 25 fs ablation laser pulse and fluence 200 mJ/cm$^2$ (Triangles and circles). TOF distribution of ablated neutral Fe from steel for 400 fs laser pulse width and 100 mJ/cm$^2$ laser fluence (squares). The red and green solid lines represent fits to a Maxwell-Boltzmann distribution. Blue line is a fit to shifted Maxwell-Boltzmann distribution. A 10-points moving average has been applied to the raw data. Error bars indicate the reproducibility of the data.

\bigskip
Fig. 4: (A) Measured TOF distribution of m/z=28 for ablation laser pulse width $\tau$= 350 and 500 fs and fluence 200 mJ/cm$^{2}$. The lines represent fits to a modified-MB distribution. For comparison, we also include the modified-MB fit of the m/z=28 TOF distribution (blue line, figure 3)(B) Surface temperature, determined from modified MBD, as a function of the laser pulse width. (C) The total yields, derived from the area of the modified MBD, as a function of the laser pulse width.

\end{document}